\documentclass[12pt,preprint]{aastex}
\usepackage{epsfig}
\usepackage{overcite}

\begin{document}
\title{Blue Straggler Stars in Galactic Open Clusters and the Simple Stellar Population Model}
\author{Y. Xin\altaffilmark{1, 2}, L. C. Deng\altaffilmark{2} and Z. W. Han\altaffilmark{1}}
\email{xinyu@bao.ac.cn}

\altaffiltext{1}{National Astronomical Observatories/Yunnan
Observatory, Chinese Academy of Sciences, PO Box 110, Kunming,
Yunnan Province, 650011, China}

\altaffiltext{2}{National Astronomical Observatories, Chinese
Academy of Sciences, Beijing, 100012, China}

\begin{abstract}

Blue straggler stars present as secure members in the Galactic open
clusters form a major challenge to the conventional picture of
evolutionary population synthesis based on the stellar evolution
theory of single stars, as illustrated in our previous work.
Expansion of our sample in the current work to include younger age
clusters provides a larger data base to expose the question raised
for the simple stellar population model. The working sample now
includes 97 Galactic open clusters of ages ranging from 0.1 to
several Gyrs. The contributions of blue straggler stars to the
integrated light of the host clusters are calculated on an
individual cluster base. A data base of observational constrained
simple stellar population model is made which has a larger age
coverage than our previous work. It is shown in this work that the
general existence of blue stragglers in star clusters of our sample
dramatically altered the predictions of convectional stellar
population model in terms of spectral energy distribution. The
integrated spectral energy distributions of the synthetic spectra of
the clusters are enhanced towards shorter wavelengths, therefore the
results of the present work will cast new lights in understanding
the properties of stellar populations.

\end{abstract}


\section{INTRODUCTION}

Although the evolutionary population synthesis (EPS) method has been
widely applied in analyzing stellar contents of local and remote
galaxies, and has been proved to be very successful in many aspects,
there are still some fundamental problems indeed in the framework of
EPS. One of the challenges to the EPS is that the effect of
complex stellar interactions which are common in all stellar
systems, such as mass exchanging and coalescence in binaries and
collisions between single and/or binaries, etc., is not considered
in the method. The conventional picture of EPS is mainly based on
the single-star evolution theory and is performed with the
evolutionary behaviors and spectra libraries of individual stars
(Bressan et al. 1994).

In order to attack such a problem, getting direct observational
spectra of simple stellar populations (SSPs) had been practiced by
former studies, such as the early work of Bica \& Alloin (1986a, b).
They undertook a direct approach using the observational integrated
spectra of star clusters, assuming that these clusters are SSPs so
that the EPS can be constrained observationally. Direct observation
of star clusters is a meaningful approach to discuss the integrated
properties of SSPs, and it is free of any assumptions about the
initial mass function (IMF) and the details of stellar evolution.
However, such a treatment can not perfectly reproduce the
conventional SSP model, since the dynamical evolution of star
clusters has modified the content of the original population, such
as gravitational evaporation of low-mass member stars. More
importantly, possible foreground star contaminations to the spectra
cannot be assessed in an ideal manner. A practical way out should be
resolving member stars photometrically using proper motion and/or
radial velocity data. By conventional SSP model through this paper,
we mean that the model is built based on single star evolution
theory.

We have carried out a plan to attack this problem semi-empirically.
The Galactic open clusters (OCs) are taken as our working sample to
explore the contributions of blue straggler stars (BSs) to the
conventional SSP model. The first set of results for old OCs with
ages greater than 1Gyr has been published (Xin \& Deng 2005,
hereafter XD05). All the member stars within a cluster are
presumably born at the same time from the same origin, therefore
should have the same age and metallicity. Star clusters have been
long regarded as one of the best objects to study the population
synthesis technique (Battinelli et al. 1994). As described in XD05,
the bulk of cluster member stars well fitted by an isochrone
represents the idea of conventional SSP model. All the other member
stars straggling away from the isochrone belong to the same
population, no matter how weird their positions in the
color-magnitude diagram (CMD) are, and have to be included when
considering the integrated light properties of the population. Among
the stragglers, BSs are of special interest because they are the
only luminous objects straggling away from the predictions of the
theory of single star evolution. It is also important to recall that
these bright objects cannot be fully understood either by current
theory of binary star evolution. Baring these in mind, the
semi-empirical model used in XD05 is still a working approach.
Besides, the nature of BSs is a challenging subject of stellar
evolution, which is under active investigations (Tian et al. 2006,
Hurley et al. 2005, Chen \& Han 2004).

Since the first identification of BSs by Sandage (1953) in the
globular cluster (GC) M3, the common existence of this kind of stars
has been proved in stellar systems of all scales and complexities
(e.g., Ahumada \& Lappaset 1995 for open clusters; Piotto et al.
2002 for globular clusters and Lee et al. 2003 for dwarf galaxies).
The typical locations of BSs in the CMD of a cluster are at the
bluer and brighter extension of the turnoff point of the host
cluster, therefore they are the most luminous and bluest hydrogen
burning stars of the cluster at the time of observation (Deng et al.
1999). The main formation mechanisms of BSs are generally related
with dynamical interactions of close binaries (Pols \& Marinus 1994)
and stellar collisions in the high-density areas (Ferraro et al.
2003). Therefore, statistical studies of BS populations in clusters
in terms of integrated light provide a clue to such stellar
interaction processes, and a reference to approaches such as binary
evolution and dynamical evolution within the cluster, and eventually
make the EPS more realistic. A larger sample is definitely needed
for this purpose.

97 Galactic OCs are studied in this paper in terms of integrated
light. A short description of the working sample is given in the
next section. In \S 3, how BSs influence the integrated properties
of conventional SSP model is discussed. The synthetic clusters'
ISEDs involving BSs contributions show significant enhancements
towards shorter wavelengths, especially in ultraviolet (UV) and blue
bands, and consequently the integrated (U-B) and (B-V) colors become
bluer. When measured with the modifications in (U-B) and (B-V)
colors, the physical parameters of OCs, especially age and
metallicity, show great uncertainties. The uncertainties are
considered in more details in \S 4. The synthetic integrated
spectral energy distributions (ISEDs) of our sample OCs are fitted
with those of the conventional SSPs of either younger ages or lower
metallicities. Based on the results of all the sample OCs, the
uncertainties in measuring the fundamental parameters of SSPs are
discussed. Finally, the concluding remarks on this work are
presented in \S 5.

\section{The Working Sample}

In XD05, relying on the definite identifications of BSs and cluster
turnoff point, we put our attention on the old Galactic OCs since
they possess sufficient number of stars for good statistics and well
populated all evolutionary stages. However, since the aim of this
series work is to detect the effect of BSs quantitatively and
systemically at ranges as large as possible in age and metallicity,
a large enough working sample is inevitably needed. The working
sample is expanded from 27 old Galactic OCs (age $\geq$ 1.0 Gyr) in
XD05 to 97 Galactic OCs in this work, including 62 intermediate (0.1
Gyr $\leq$ age $<$ 1 Gyr) and 35 old (age $\geq$ 1 Gyr) clusters.

The basic parameters of the sample OCs are given in Table
\ref{tab1}, where column 1 is cluster name; columns 2-4 give the age
($\log (\rm{age})$), color excess (E(B-V)) and distance modulus
(DM); columns 5-6 give the metallicity (Z) and the [Fe/H] values;
columns 7-8 are the N$_2$ (number of stars within two magnitudes
below the turn-off) and N$_{BS}$ (number of BSs) numbers of the
sample clusters. References for parameter adopted in this work are
listed in column 9.

In order to keep the criterion homogeneous for selecting BSs in the
clusters, we use exclusively the photometric data from Ahumada \&
Lappaset's BS catalog (1995, hereafter AL95). N$_2$ and N$_{BS}$
numbers are also from the same source. N$_2$ is defined as the
number of stars within two magnitudes interval below the turnoff
point of the host cluster. The detailed explanation of N$_2$ number
can be found in XD05. The physical parameters of sample clusters,
including age, E(B-V), Z and DM, are mainly extracted from more
recent literatures, the latest photometric data of the OCs are then
used in this work in order to get optimistic parameters of the OCs
and to select the BSs candidates in each cluster based on the
observed CMDs more reliably. For sample OCs without accurate
photometric data, the physical parameters are chosen in such a way
that locations of majority of BSs are reasonable in the CMDs.
Detailed derivation for every parameter is listed in Table
\ref{tab1}.

The metallicity of a cluster is usually given as [Fe/H] based on
spectral observation and abundance analysis, but we need the
information in Z to build conventional SSPs. In such circumstance,
an empirical relation between Z and [Fe/H] (Carraro et al. 1994) is
adopted in the work,
\begin{equation}
\log Z=1.03\times[Fe/H]-1.698.
\end{equation}

When metallicity Z is marked as ``0.02?'' for a cluster in Table 1,
it means that there is nothing available on metallicity for that
cluster. In this case, we assume solar metallicity (Z=0.02) for it.
The Galactic OCs generally possess younger ages and locate closer to
the Galactic plane when compared with the Galactic GCs, which means
that, at the first approximation, the assumption of Z=0.02 will not
cause any essential errors for clusters considered in this work.

\begin{figure}
\plotone{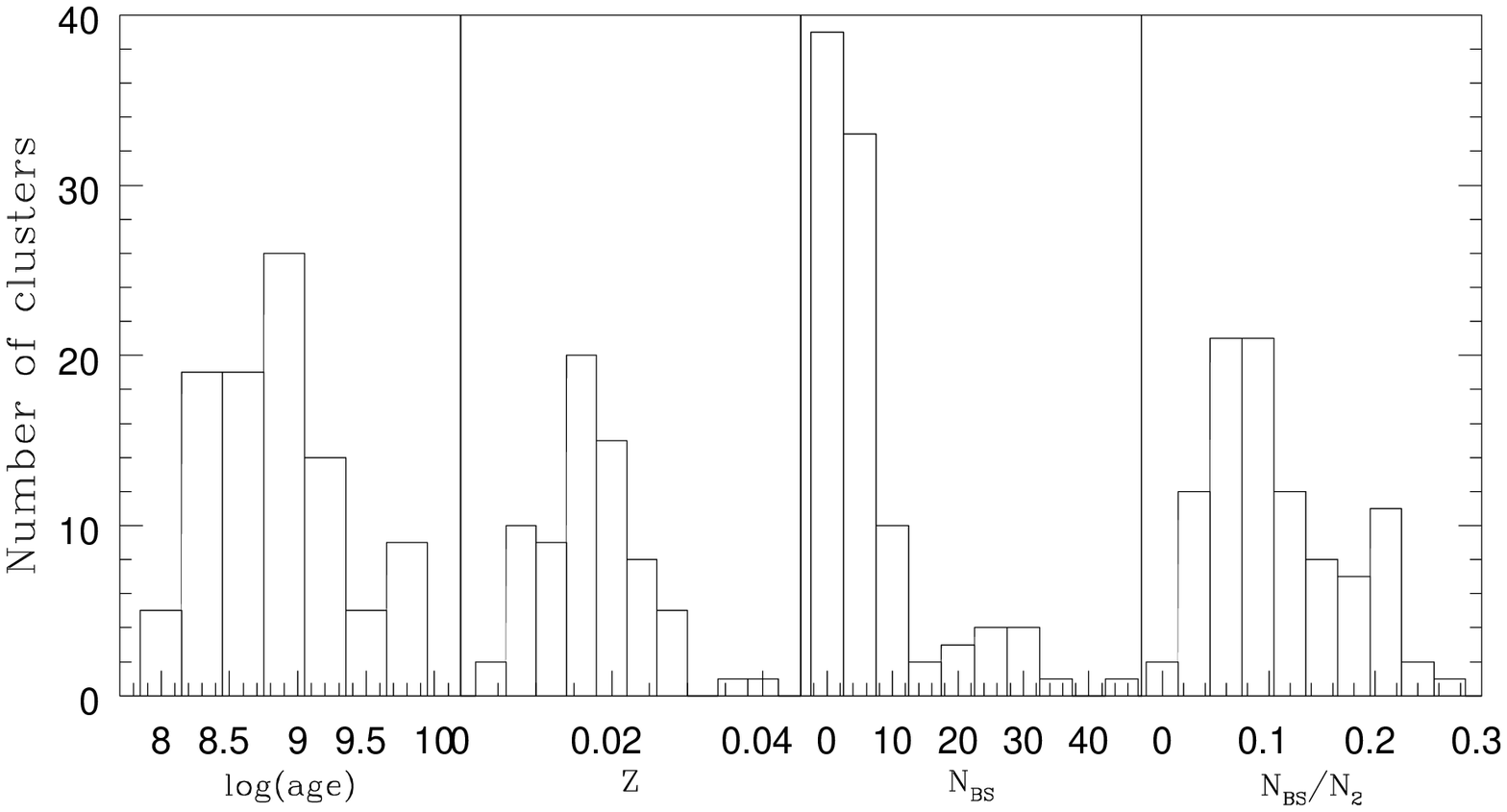} \caption{Number distributions of the
sample OCs according to four different parameters: $\log
(\rm{age})$, Z, N$_{BS}$ and N$_{BS}$/N$_2$.\label{fig1}}
\end{figure}

\begin{figure}
\plotone{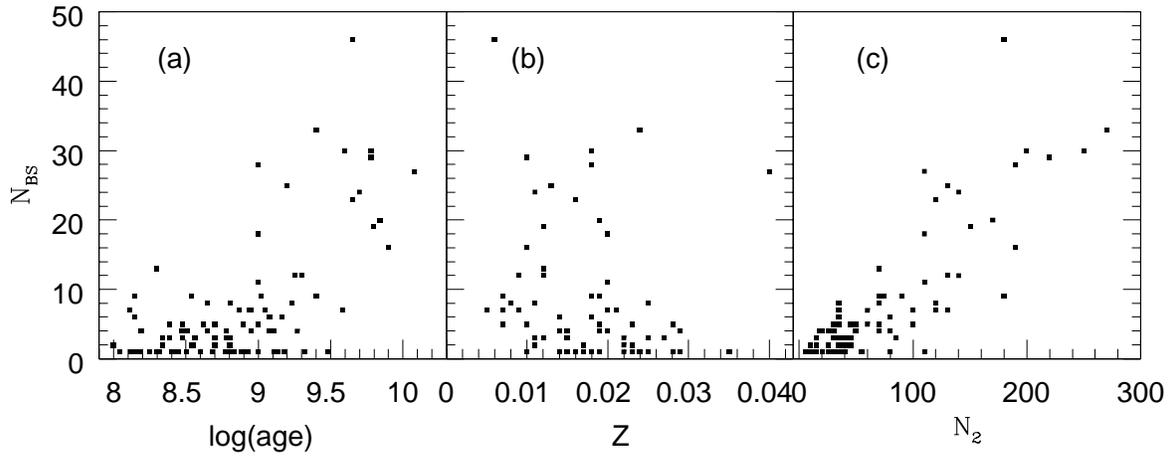} \caption{N$_{BS}$ distributions
according to three different parameters: $\log (\rm{age})$, Z and
N$_2$. \label{fig2}}
\end{figure}

Figure \ref{fig1} shows the number distributions of total 97
selected OCs as functions of four different parameters: $\log
(\rm{age})$, Z, N$_{BS}$ and N$_{BS}$/N$_2$. Only the clusters with
published metallicities are used to plot the distribution in Z. As
shown in the figure, most of our sample OCs possess Z values around
Z=0.02, pretty much like what we expected, and hence it is an
observational support to our assumption of Z=0.02 for the sample OCs
without accurate metallicity determinations. The ratio of
N$_{BS}$/N$_2$ can be regarded as a specific BS frequency in a
cluster. It can be used as a probe for the cluster internal dynamic
processes concerning BSs formation. The three panels in Figure
\ref{fig2} show the correlations between N$_{BS}$ and three
different parameters: $\log (\rm{age})$ (Figure~\ref{fig2}a), Z
(Figure \ref{fig2}b) and N$_2$ (Figure \ref{fig2}c), where there is
no indication of obvious correlations between N$_{BS}$ and $\log
(\rm{age})$, Z, even in this larger sample compared with the
previous result in XD05. While a well defined correlation between
N$_{BS}$ and N$_2$ (Figure \ref{fig2}c) is clearly shown, which
infers that BSs formation is quite similar in all OCs, and is
positively correlated with the total number of stars. The BS content
in a cluster is almost independent either of age or of metallicity
of the host cluster.

\section{MODIFICATION TO THE CONVENTIONAL SSP MODEL}

As described in XD05, the stellar population corresponding to a
cluster is assumed to be composed of two components. One accounts
all member stars that are well fitted by an isochrone of single star
evolution theory, i.e. a conventional SSP model. It is named as
``SSP component'' and its ISED can be directly built from
theoretical model. The other component should include all the other
member stars straggling away from that isochrone, and our attention
is put only on BSs because they are bright. The other stragglers are
not quite important in terms of contributions to the total light:
the red stragglers in the middle of CMD are usually rare and less
luminous; the underlopers and stars bluer than the main sequence but
well below the turnoff are too faint to be important. The BSs fixed
by photometric observations form the ``BS component''. More detailed
descriptions of model construction can be found in XD05.

In this work, the ISEDs of SSP components are extracted from the
low-resolution conventional SSP model of Bruzual \& Charlot (2003,
hereafter BC03). The quoted ISEDs are built based on Padova
isochrone (Bertelli et al. 1994), Salpeter IMF (Salpeter 1955) and
Lejeune stellar spectra library (Lejeune et al. 1997). For details
we refer BC03. The spectra of BS components are also made based on
the Lejeune spectra library following a process of interpolation and
fitting, see XD05 for details. Summing up the ISEDs of these two
components, we get the synthetic ISEDs of the clusters. The
differences between the ISED of the SSP component and the synthetic
ISED of a cluster are exactly the BSs contributions. In this work,
we are going to calculate such contributions for all the clusters in
our sample in terms of both ISEDs and broadband colors.

\subsection{ISED of Conventional SSP Model}

\begin{figure}
\begin{center}
\plotone{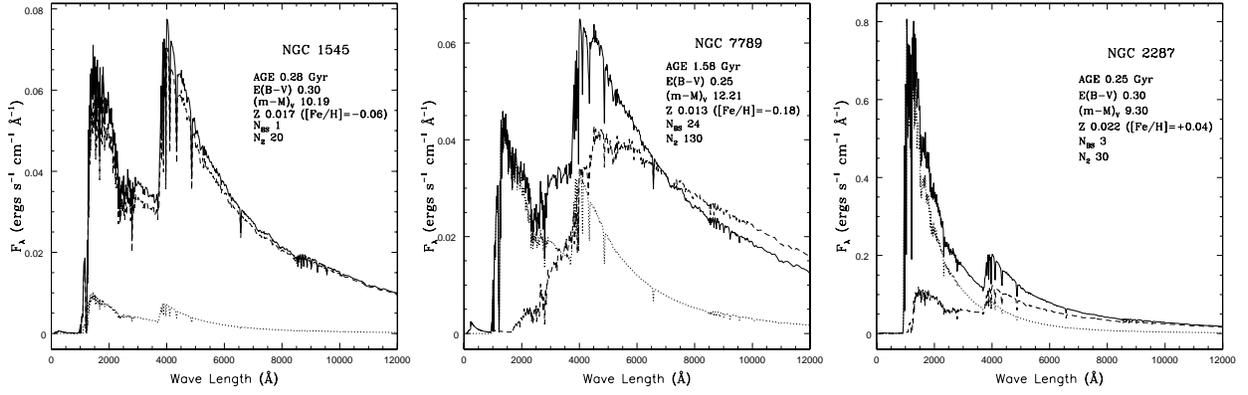}
\caption{ISED modifications. $\it{Dotted}$ $\it{line:}$ ISED of the
BS component. $\it{Dash}$ $\it{line:}$ ISED of the SSP component.
$\it{Solid}$ $\it{line:}$ Synthetic ISED of the cluster.
\label{fig3}}
\end{center}
\end{figure}

In order to demonstrate the effect of BSs on the ISEDs of the
conventional SSPs represented by OCs, the ISEDs of three clusters
are given in Figure \ref{fig3}. The abscissa is the wavelength in
angstrom, the ordinate is the integrated flux. The dotted line shows
the ISED of BS component, the dash line shows that of the SSP
component, and the solid line is the synthetic ISED of a cluster. It
is clearly seen in Figure \ref{fig3} that the ISEDs of the
conventional SSPs have all been modified by BS components to some
extent for all the clusters. In the left panel in Figure \ref{fig3},
normal stars still dominate, which means that the cluster's ISED is
determined by regular stars in that situation; while in the right
panel, the synthetic ISED of the cluster is completely overwhelmed
by the BS component.

The synthetic ISEDs show systematic enhancements towards shorter
wavelengths due to addition of hotter stars (BSs) in the population,
especially in UV and blue bands, which makes the ISEDs hotter than
those of the conventional SSP model. The amplitudes of such
modifications due to BS components depend on the detailed physical
properties of both BSs and the host clusters, i.e., membership
measurements, BSs positions in the CMDs, spatial configuration of
the host clusters and the star richness of the clusters. All these
things are crucial for the understanding of the real stellar
populations containing stars with peculiar behaviors, like BSs
discussed in our work. Detailed discussions on the scenario of BS
property and its consequence on ISED of a population can be found in
XD05.

\subsection{Broadband color of conventional SSP model}

Photometric observations, usually in broadband colors, are more
frequently used than spectrophotemetry when trying to understand
stellar populations in remote galaxies in unresolvable conditions.
Including of BS components in the conventional SSP model, the ISEDs
are inclined to the UV and blue bands and consequently make the
integrated (U-B) and (B-V) colors bluer. The use of these two colors
is practically a good choice to show the photometric effect of BSs.
This fact provides a quantitative way to measure the modification to
the conventional SSP model. Bluer broadband colors in real stellar
populations with respect to the conventional model at the same
parameter can be misunderstood by EPS method with either younger
ages or lower metallicities. Therefore, (U-B) and (B-V) colors are
taken as detectors in this work to quantify BSs contributions to the
conventional SSP model. The colors are obtained by convolving the
ISEDs with corresponding filter responses.

\begin{figure}
\plotone{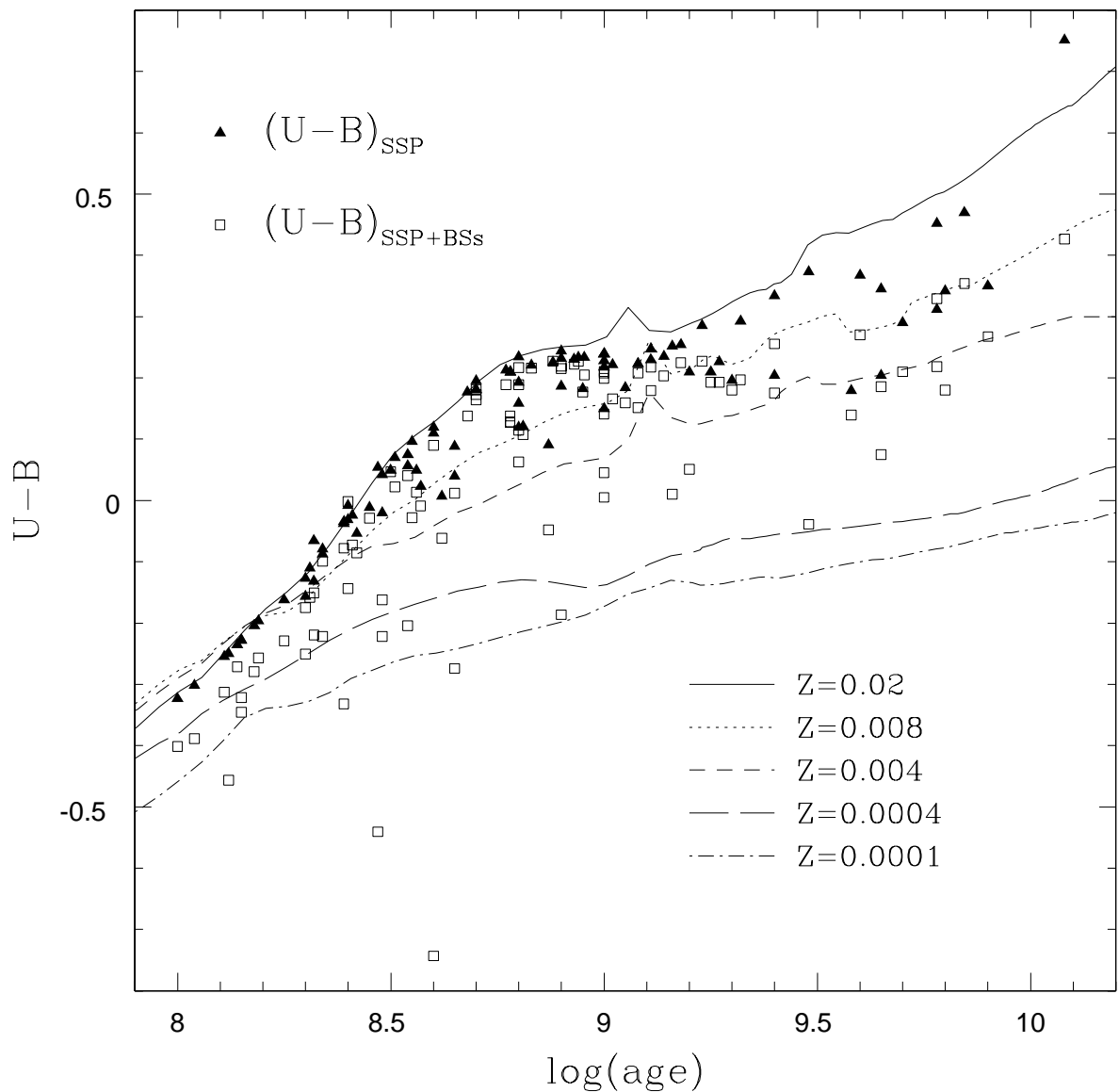} \caption{(U-B) colors modified by BSs.
The $\it{Solid}$ $\it{triangles}$ are (U-B) colors of the SSP
components. The $\it{Open}$ $\it{squares}$ are (U-B) colors of our
sample OCs including BS contributions. Five lines in different types
are the theoretical (U-B) colors from BC03 of different
metallicities. \label{fig4}}
\end{figure}

\begin{figure}
\plotone{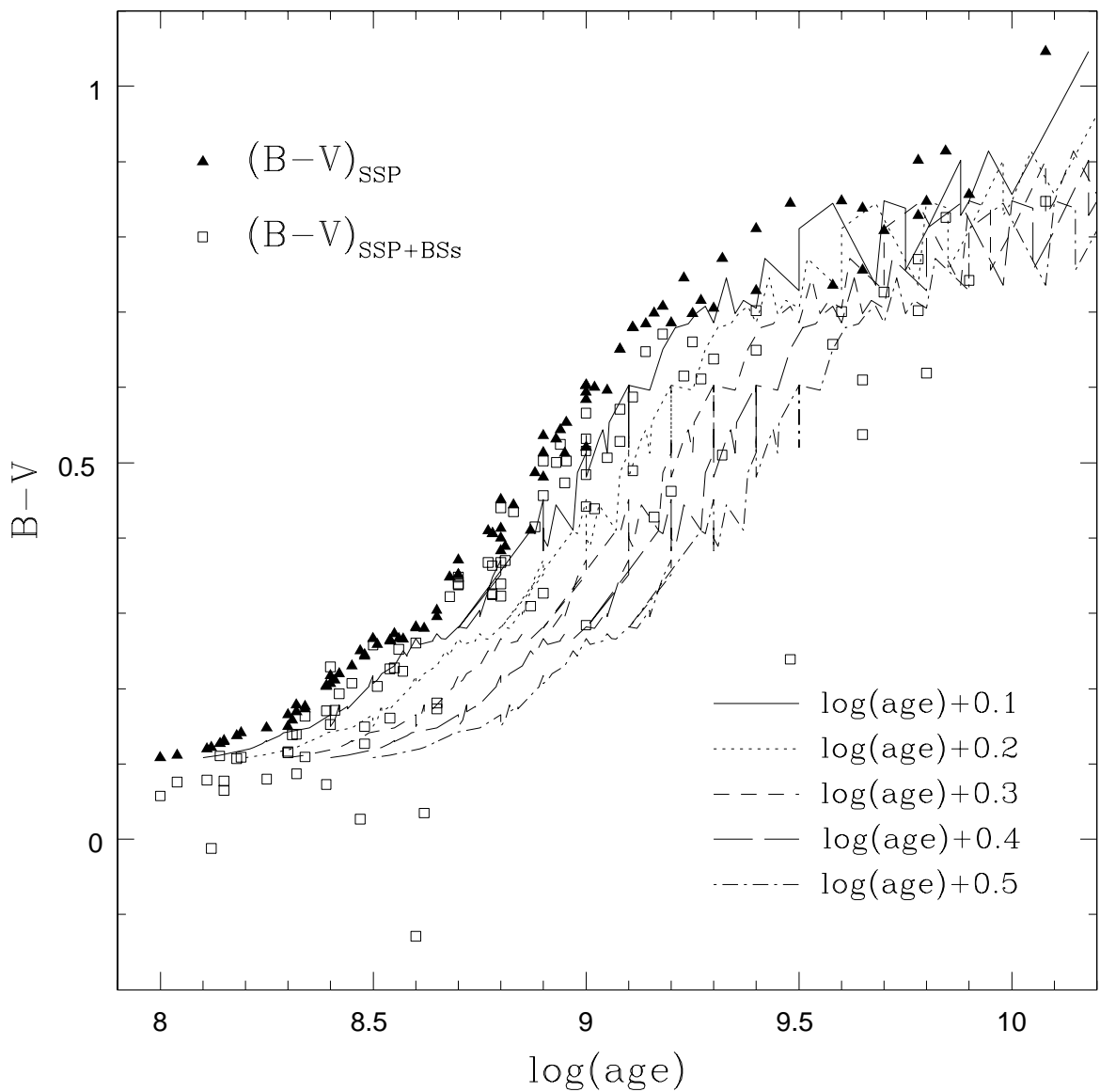} \caption{(B-V) colors modified by
BSs. The $\it{Solid}$ $\it{triangles}$ are (B-V) colors of the SSP
components. The $\it{Open}$ $\it{squares}$ are (B-V) colors of our
sample OCs including BS contributions. Five lines in different types
are plotted by keeping the theoretical colors the same while
expanding the ages, see text in details. \label{fig5}}
\end{figure}

As shown in Figure \ref{fig4} and Figure \ref{fig5}, respectively,
the (U-B) and (B-V) colors have been dramatically modified by BSs.
These two figures are plotted using all the 97 sample OCs, including
the clusters without accurate metallcity measurements and assumed
Z=0.02. The abscissa is the logarithmic ages of sample OCs. The
ordinate is the broadband colors as indicated. The solid triangles
are colors of the SSP components, and the open squares are colors of
our sample OCs involving BSs contributions. In order to interpret BS
influence on SSP in terms of color, we put a set of theoretical
colors of SSP model as functions of age by lines in different types
in each figure. In Figure \ref{fig4}, the five lines demonstrating
the convectional SSP model in (U-B) are from BC03 in low-resolution
case with different metallicities. In Figure \ref{fig5}, the five
lines present the same fact as in figure \ref{fig4} for (B-V) color
with expanded ages so that uncertainty in age can be measured. The
lines are given by keeping the invariable (B-V) color of SSP
components (solid triangles) while expanding the ages for given
values. For example, when keeping the same (B-V) color of each
triangle point but changing the corresponding age by 0.1 dex larger
than the original one, the solid line of ``$\log (\rm{age})$+0.1''
is plotted. Similarly, other lines are made with different age
increments. Detailed number for each line is given in the figure.

As shown in Figure~\ref{fig4}, the (U-B) colors of SSP components of
younger ages ($<$ 1.0 Gyr) stay close the theoretical line of BC03
of Z=0.02, while those of old ages ($\geq$ 1.0 Gyr) have
considerable deviations from that line. Such a pattern is not
understood with the distribution in the Galactic plane and the
history of formation and evolution of the clusters. However, it
infers that for younger clusters, solar value is a good
approximation if no accurate determination for metallicity is
available. For the old clusters in the sample, the determination of
metallicity is more reliable, therefore the spread may be due to
true scatter in metallicity.

In contrast to the SSP components, the (U-B) and (B-V) colors of the
clusters including BSs contributions scatter much larger in both age
and metallicity plots (open squares in Figures~\ref{fig4} and
\ref{fig5}). It is apparent that large differences will be expected
when trying to get age and metallicity estimations for the clusters
from the figures using the conventional SSP model. The theoretical
colors described as different lines in these two figures cover the
metallcity region from Z=0.0001 to Z=0.02 and the age region from
``$\log (\rm{age})$+0.1'' to ``$\log (\rm{age})$+0.5'', but even
such big enlargements of age or metallcity parameter can not
completely interpret all the modified colors. It means that when
trying to get the basic parameters of real stellar populations using
broadband colors, the conventional SSP model can make the results
very weird as we demonstrate for these OCs sample. Therefore we
suggest that, in order to understand correctly stellar contents in
various stellar systems, the results of stellar interactions,
especially the bright BSs, should be implemented into the SSP model
and EPS framework in a systematic way.

\section{UNCERTAINTIES OF THE CURRENT CONVENTIONAL SSP MODEL}

In previous section, it is shown that the ISEDs and broadband (U-B)
and (B-V) colors of our sample OCs are dramatically modified by
taking into account the BS populations. The modifications can
introduce sizable uncertainties on the basic physical parameters of
a stellar population if one keeps using the convectional SSP model.
Taking a few clusters as example, it is possible to quantify these
uncertainties. Focus is put on the uncertainties in age and
metallicity in fitting the cluster's ISED using that of conventional
SSP model.

\subsection{Fitting the synthetic ISEDs of OCs}

One of the main effect of BSs is to make the synthetic ISEDs of our
sample OCs hotter, i.e. the spectra get enhanced in UV and blue
bands. The synthetic hotter (than the fitted isochrone can tell)
ISEDs can still be well fitted with an SSP model of either younger
age or lower metallicity than what is read from isochrone fitting in
the CMD. Based on this, the synthetic ISEDs of our sample OCs are
regarded as the real (observational) ones and are to be fitted by
ISEDs of conventional SSP model with depressed ages or
metallicities, and then the differences between cluster real
parameters and fitting results can be used to discuss the fitting
uncertainties that are potentially existing in all applications of
the EPS scheme.

\begin{figure}
\plottwo{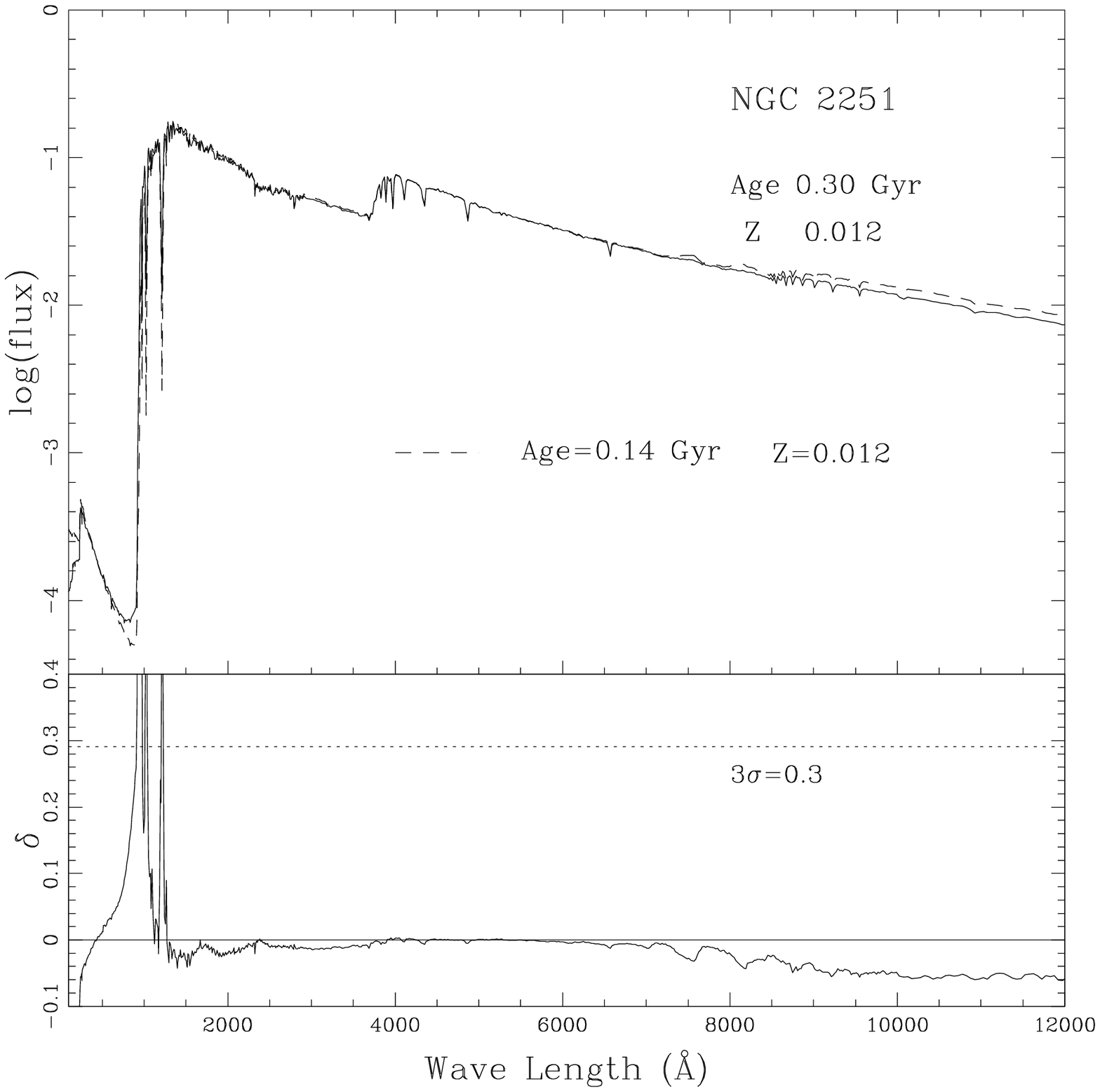}{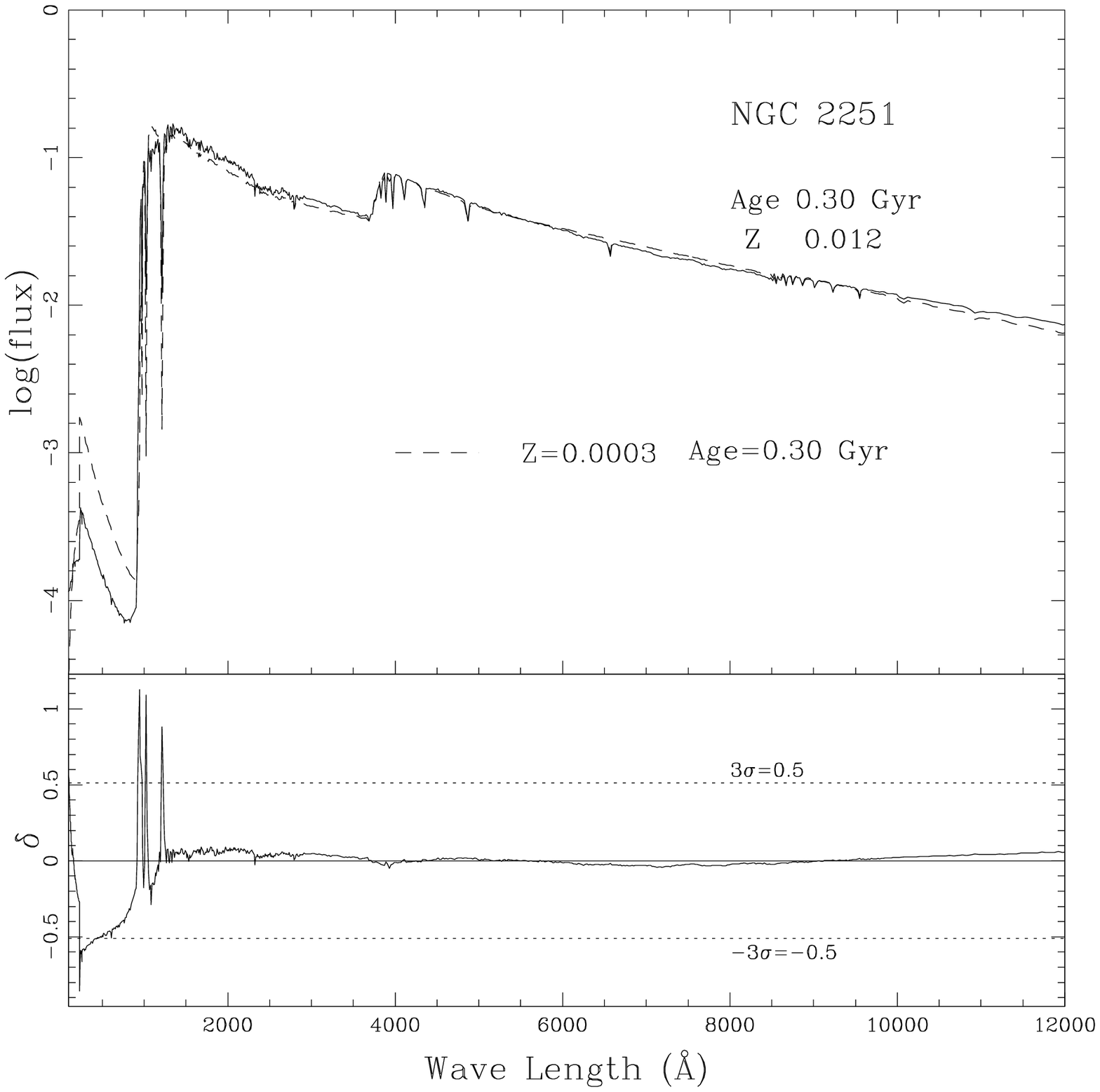}
\caption{Fitting the synthetic ISED of NGC 2251 with the
conventional SSP model. The abscissa is the wavelength in angstrom.
The ordinate is the logarithmic value of cluster ISED normalized at
5500 {\AA}. $\delta$ is the differences between the synthetic and
the conventional ISEDs. Standard deviation $\sigma$ of $\delta$ is
given in $\it{dotted}$ $\it{lines}$ in $\pm$3$\sigma$. The left
panel is the fitting result of keeping the same metallicity but
depressing age. The right panel is the opposite, keeping the same
age but lowering metallicity. The synthetic ISED is plotted in
$\it{solid}$ $\it{line}$. The conventional ISED is given in
$\it{dash}$ $\it{line}$. The real parameters of the cluster are
listed in the top right corner in each panel. The parameters of the
theoretical ISED are given below the fitting line.\label{fig6}}
\end{figure}

Taking NGC 2251 as an example, the best-fitting results are
presented in Figure \ref{fig6}. In the two panels, the abscissa is
the wavelength in angstrom. The ordinate is the logarithmic value of
the absolute flux of ISED normalized at wavelength of 5500 {\AA}. In
the lower frames, the fitting residual $\delta$ between the
synthetic ISED of the cluster and the best-fitting ISED of
conventional SSP model is given, together with the standard
deviation $\sigma$ in the region of $\pm$3$\sigma$ in dotted lines.
The left panel presents the best-fitting result with metallicity
untouched, but depressing the age parameter, while the right panel
is that just the opposite, keeping the same age but depressing the
metallicity. The synthetic ISED of the cluster is plotted in solid
line. The ISED of conventional SSP model is shown by the dash line.
As a comparison, the real parameters of the cluster are given in the
top right corner in each panel. The parameters of the fitting model
ISED are labeled below the fitted dash line.

As shown in Figure \ref{fig6}, if we let the parameter of age or
metallicity free, conventional SSP model can fit perfectly most of
the features of the synthetic ISED of the cluster, especially the
whole optical region. In this case, the broadband photometry tells
no difference. The main exception is focused on the extreme UV band,
where the availability of observation and the accuracy of the
theoretical model are both not perfect. That is to say, the
conventional SSP model is certain to do good job on fitting the
observational ISEDs of stellar populations, but the fitting results
indeed suffer great uncertainties, as shown by this example.

\subsection{Fitting uncertainties in age}

By doing the same fitting work for all the 97 sample OCs as what is
done for NGC 2251 (Figure \ref{fig6}), the parameter differences
between real cluster parameters (derived directly from fitting the
resolved stars in CMD) and the fitting results (regarded as
photometric measurements in unresolvable conditions) can then be
collected to achieve the fitting uncertainties analysis.

\begin{figure}
\plotone{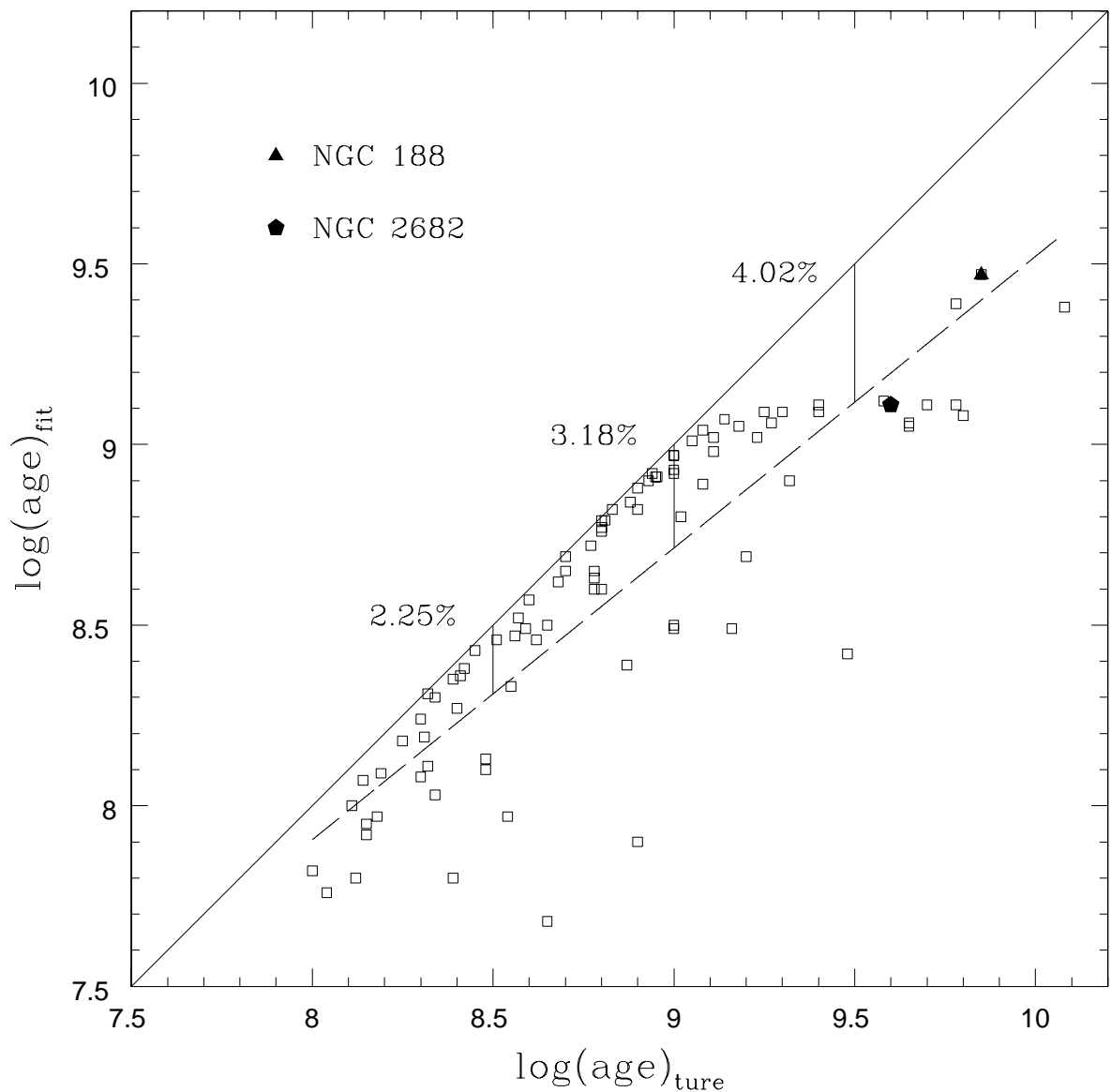} \caption{Fitting uncertainty in age.
The abscissa is the real age of the cluster. The ordinate is the
fitting age by the conventional SSP model. $\it{Open}$
$\it{squares}$ present age differences with and without the BS
contributions. $\it{Dash}$ $\it{line}$ is the least square fitting
results of all the $\it{open}$ $\it{squares}$. Results for NGC 188
and NGC 2682 (M67) are marked by $\it{solid}$ $\it{triangle}$ and
$\it{solid}$ $\it{pentagon}$, respectively. \label{fig7}}
\end{figure}

Figure \ref{fig7} demonstrates the age uncertainties. The abscissa
is the logarithmic age values of our sample clusters from the best
isochrone fitting results on the CMDs of the clusters. This is the
most reliable age measurement and these ages can be regarded as the
true ones, thus the abscissa is marked by $\log (\rm{age})_{true}$.
The ordinate is the age values made by fitting the synthetic ISEDs
of our sample OCs with the ISEDs of conventional SSP model. It is
marked by $\log (\rm{age})_{fit}$. If these two ages of clusters
were the same, all the points would stay in the diagonal line. The
open squares show the age determinations by taken into account the
BS components.

As shown in Figure \ref{fig7}, almost all the open squares locate
below the diagonal, which means that when fitting the observed ISEDs
of stellar populations (the synthetic ISED of a cluster) with the
conventional SSP model, the ages will be systematically younger than
the true ages as isochrone fitting from the photometric data. The
dash line in Figure \ref{fig7} shows the least square fitting
results of all the open squares, which marks the uncertainties on
average as the conventional SSP model can make in unresolved
conditions. If we evaluate the underestimation level by the ratio of
($\log (\rm{age})_{true}$ - $\log (\rm{age})_{fit}$)/$\log
(\rm{age})_{true}$, the logarithmic age value underestimation goes
from about 2\% at $\log (\rm{age})$=8.5, to 3\% at $\log
(\rm{age})$=9 and 4\% at $\log (\rm{age})$=9.5 as shown in
figure~\ref{fig7}, becoming larger for older ages. The two well
studied clusters, NGC 188 and NGC 2682 (M67), are marked in the plot
with solid triangle and pentagon, respectively. As the BS
populations are reliably know in them, their positions in the figure
can be thought as calibrations of our method. Indeed, they agree
quite well with our analysis.

\subsection{Fitting uncertainties in metallicity}

\begin{figure}
\plotone{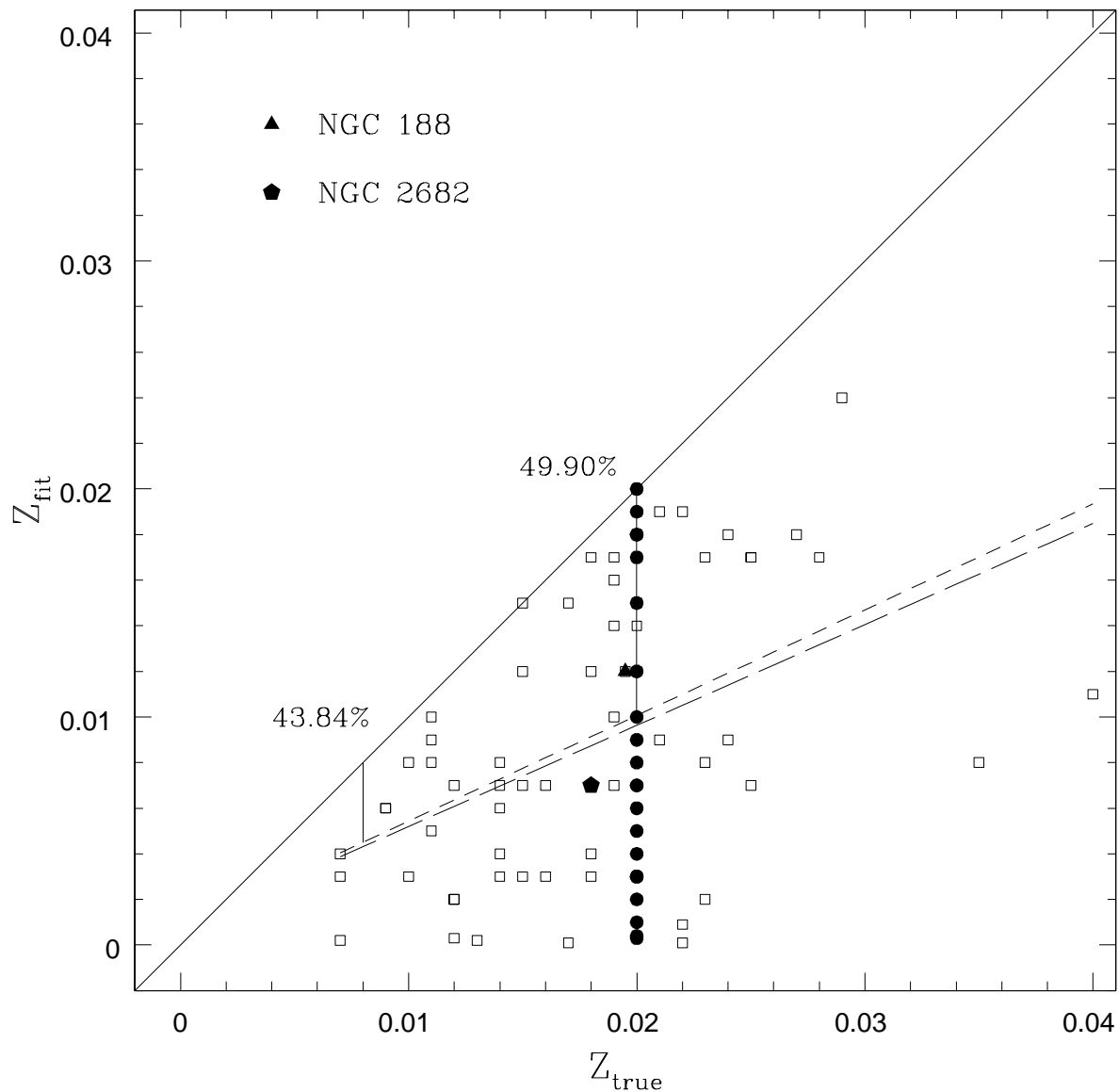}\caption{Fitting uncertainty in
metallicity. The abscissa is the real metallicity of the cluster.
The ordinate is the fitting metallicity by the conventional SSP
model. $\it{Open}$ $\it{squares}$ present differences in metallicity
with and without the BS contributions. $\it{Solid}$ $\it{circles}$
are results of those OCs assuming Z=0.02. $\it{Short}$ $\it{dash}$
$\it{line}$ is the least square fitting results of all the
$\it{Open}$ $\it{squares}$. $\it{Long}$ $\it{Dash}$ $\it{line}$ is
the least square fitting results of all points. Results for NGC 188
and NGC 2682 (M67) are marked by $\it{solid}$ $\it{triangle}$ and
$\it{solid}$ $\it{pentagon}$, respectively. \label{fig8}}
\end{figure}

Figure \ref{fig8} is for the analysis of metallicity uncertainty.
The clusters with their BS contribution overwhelming are not used
because the fitting results are obviously not realistic. On the
other hand, the BSs in these excluded clusters need more photometry
and membership studies. The total number of 83 clusters are plotted
in Figure~\ref{fig8}. The abscissa (Z$_{true}$) is the true
metallicities of sample OCs, also fixed by isochrone fitting
technique using resolved photometric data. The ordinate (Z$_{fit}$)
is the metallicity values made by fitting the synthetic ISEDs with
those of conventional SSP model of lower metallicities. The open
squares present metallicity values of two measurements. The solid
circles are the results of the clusters without metallicity
references and assumed Z=0.02 (entries in Table \ref{tab1} marked
``Z=0.02?''). The short-dash line is again a least square fitting to
data points with metallicity measurements, and the long-dash line is
the least square fitting results of all the points (including solid
circles). The short-dash line with metallicity measurements is used
for the discussion below.

It is clearly shown in Figure \ref{fig8} that vast majority of data
points locate below the diagonal. In a similar way as for age
analysis, the ratio of (Z$_{true}$-Z$_{fit}$)/Z$_{true}$ is used to
measure the uncertainty level, which goes from nearly 44\% at
Z=0.008 to 50\% at Z=0.02. Again, NGC 188 and NGC 2682 (M67) are put
in the plot as references.

Compared with isochrone fitting technique, spectra observation of
individual stars is no doubt a more reliable way to get metallicity
information of a star cluster. However, it still suffers some
problems. The accuracy is connected with observational conditions,
extinction and contamination effect. The results coming from that
are also not perfectly ensured since the metallicity determination
based on different stars in different regions in a cluster will
produce different results, and due to selection effect, red giant
branch stars in a cluster are easier taken as observational
candidates than main sequence stars. Besides observational
uncertainties, theoretical model is another uncertainty source for
this statistic results, since there are originally only limited
metallicity grids from both theoretical evolutionary tracks (Bressan
et al. 1994) and model atmospheres (Lejeune et al. 1997). Stellar
spectra used for synthetic ISEDs of clusters rely heavily on
interpolation. Therefore, the metallicity uncertainties presented in
Figure \ref{fig8} should be considered as qualitative statistic
results. More significant movements on this problem should be
supported by more reliable observational data and more detailed
model calculations.

\section{CONCLUSIONS}

As a follow-up work of XD05, the main aim of this work is to extend
the study to larger sample and to quantify the influence of BSs on
the conventional SSP model. With a sample of 97 Galactic OCs of ages
$\geq$ 0.1 Gyr, we confirm the results of XD05 that the integrated
spectral properties OCs are dramatically modified by the BS contents
in the clusters. The ISEDs are greatly enhanced towards shorter
wavelength, becoming significantly bluer, and consequently the
broadband (U-B) and (B-V) colors are modified accordingly. When the
conventional SSP model is adopted in understanding stellar contents
in galaxies, either using spectra or broadband colors, great
uncertainties in age and/or metallicity will be brought in.

Directly fitting the synthetic ISEDs of our sample OCs with the
ISEDs of conventional SSP model can still result in nice fittings,
if one lets the parameter of age or metallicity free. But the
fitting parameters have significant underestimations when compared
with the true parameters made photometrically. In the parameter
regions covered by our sample OCs, the logarithmic age value
underestimation rates ranges 2\% at $\log (\rm{age})$=8.5, 3\% at
$\log (\rm{age})$=9.0 and 4\% at $\log (\rm{age})$=9.5. There is a
trend for such a ratio to become larger for greater ages.
Also due to the presence of BSs in real stellar populations, the
metallicity parameter can also be seriously underestimated by
conventional SSP models. This is true at least qualitatively within
the scope of current work due to the uncertainties in measuring the
metallicity independently by other means.

Taking OCs as general representatives of SSPs in galaxies,
and limiting our results to the age and metallicity ranges covered
by the working sample of current work,
the conventional SSP model derived from the
single star evolution theory is seriously altered by the descendants
of stellar interactions that are presumably common in normal
environments. BSs discussed in our work are the most important
components that are proved to be widely existing in all stellar
systems (Stryker 1993). Therefore, the effect of BSs must be taken
into account when EPS technique is applied to study the properties
of stellar populations in galaxies.

Theoretical refinery of the conventional SSP model still needs more
studies. Thorough investigations of several issues should be put
forward, that including: the individual processes of stellar
interactions including binary and multiple systems and the overall
consequences of all these interactions to the population
statistically, and effects of dynamical evolution of the stellar
systems. Before such SSP model eventually become available,
emprically corrected models would be a good practice. Based no
current knowledge of interactive binary evolution and collisional
processes, and the observational cluster data base, such models can
be made. Work to this purpose is now in progress.

\acknowledgements We would like to thank the Chinese National
Science Foundation for support through grants 10573022, 10333060,
10521001 and 10433030.

\begin{deluxetable}{lrrrlrrrl}
\tabletypesize{\small} \tablewidth{15cm}\tablecaption{Parameters of
the Sample Clusters \label{tab1}} \tablehead{ \colhead{Cluster Name}
& \colhead{$\log (\rm{age})$} & \colhead{E(B-V)} & \colhead{DM} &
\colhead{Z} & \colhead{[Fe/H]} & \colhead{N$_2$} &
\colhead{N$_{BS}$} & \colhead{Ref.}}

\startdata
 Berkeley 11  &  8.04  & 0.95  &  11.71  & 0.02?  & \nodata &   15 &  1  & 1,2  \\
 Berkeley 32  &  9.80  & 0.08  &  12.60  & 0.012  &  -0.20  &  150 & 19  & 2,3  \\
 Berkeley 39  &  9.78  & 0.11  &  13.60  &  0.01  &  -0.31  &  220 & 29  & 2,4,5  \\
 Berkeley 42  &  9.32  & 0.76  &  11.31  & 0.02?  & \nodata &   20 &  1  & 1,2 \\
    Blanco 1  &  8.32  & 0.01  &   7.18  & 0.035  &   0.23  &   10 &  1  & 1,2,6 \\
Collinder 223 &  8.00  & 0.25  &  13.00  & 0.02?  & \nodata &   25 &  2  & 2,7 \\
     Hyades   &  8.80  & 0.01  &   3.39  & 0.029  &   0.15  &   40 &  1  & 2,8 \\
      IC 166  &  9.00  & 0.80  &  15.65  &  0.02  &   0.00  &  110 & 11  & 2,9,10 \\
     IC 1311  &  8.95  & 0.45  &  14.10  &  0.02  &   0.00  &  100 &  7  & 2,11 \\
     IC 1369  &  9.16  & 0.57  &  11.59  & 0.02?  & \nodata &   35 &  6  & 1,2,8  \\
     IC 2488  &  8.25  & 0.24  &  11.20  & 0.019  &  -0.02  &   10 &  1  & 2,12 \\
     IC 2714  &  8.50  & 0.36  &  11.68  & 0.015  &  -0.12  &   80 &  1  & 2,13 \\
     IC 4651  &  9.23  & 0.10  &  10.03  & 0.025  &   0.10  &   35 &  8  & 2,14 \\
     IC 4756  &  8.90  & 0.23  &   7.60  & 0.022  &   0.04  &   55 &  1  & 2,15 \\
      King 2  &  9.78  & 0.31  &  13.80  & 0.02?  & \nodata &  250 & 30  & 1,2 \\
      King 8  &  8.62  & 0.58  &  14.03  & 0.007  &  -0.46  &   35 &  5  & 1,2 \\
     King 11  &  9.70  & 1.00  &  11.70  & 0.011  &  -0.27  &  140 & 24  & 1,2,16 \\
  Melotte 66  &  9.65  & 0.16  &  13.75  & 0.006  &  -0.53  &  180 & 46  & 2,17 \\
 Melotte 105  &  8.40  & 0.52  &  11.80  & 0.02?  & \nodata &   25 &  1  & 2,18 \\
 Melotte 111  &  8.60  & 0.00  &   4.77  & 0.019  &  -0.03  &   10 &  1  & 2,19 \\
     NGC 188  &  9.85  & 0.08  &  11.35  & 0.019  &  -0.01  &  170 & 20  & 2,20 \\
     NGC 381  &  8.77  & 0.34  &  11.18  & 0.024  &   0.07  &   25 &  1  & 1,2,6 \\
     NGC 752  &  9.14  & 0.04  &   8.43  & 0.016  &  -0.09  &   25 &  1  & 1,2,6 \\
    NGC 1027  &  8.55  & 0.33  &  10.46  & 0.023  &   0.06  &   40 &  2  & 2,6,8 \\
    NGC 1193  &  9.90  & 0.12  &  13.80  &  0.01  &  -0.29  &  190 & 16  & 2,21 \\
    NGC 1245  &  9.02  & 0.29  &  12.27  & 0.018  &  -0.05  &   75 &  9  & 2,22,23 \\
    NGC 1252  &  9.48  & 0.02  &   9.04  & 0.02?  & \nodata &    7 &  1  & 1,2 \\
    NGC 1342  &  8.65  & 0.32  &  10.11  & 0.014  &  -0.16  &   20 &  4  & 1,2,6 \\
    NGC 1528  &  8.57  & 0.30  &  10.19  & 0.011  &  -0.27  &   35 &  3  & 2,6,8 \\
    NGC 1545  &  8.45  & 0.30  &  10.19  & 0.017  &  -0.06  &   20 &  1  & 1,2,6 \\
    NGC 1664  &  8.47  & 0.25  &  11.17  & 0.029  &   0.15  &   30 &  4  & 1,2,6,8 \\
    NGC 1778  &  8.18  & 0.23  &  11.53  & 0.02?  & \nodata &   15 &  1  & 2,6 \\
    NGC 1817  &  9.05  & 0.19  &  10.90  & 0.009  &  -0.34  &   35 &  7  & 2,24 \\
    NGC 1901  &  8.93  & 0.02  &   8.12  & 0.02?  & \nodata &   10 &  1  & 2,6 \\
    NGC 1912  &  8.56  & 0.25  &  10.91  & 0.015  &  -0.11  &   45 &  3  & 2,6,25 \\
    NGC 2099  &  8.81  & 0.23  &  11.50  & 0.011  &  -0.25  &  120 &  8  & 2,26 \\
    NGC 2168  &  8.30  & 0.255 &   9.60  & 0.012  &  -0.21  &   70 & 13  & 2,27,28 \\
    NGC 2204  &  9.40  & 0.08  &  13.10  & 0.007  &  -0.44  &  180 &  9  & 2,29 \\
    NGC 2236  &  8.54  & 0.48  &  12.33  & 0.017  &  -0.07  &   10 &  2  & 1,2,8 \\
    NGC 2243  &  9.58  & 0.055 &  13.15  & 0.005  &  -0.57  &  120 &  7  & 2,30 \\
    NGC 2251  &  8.48  & 0.21  &  11.21  & 0.012  &  -0.20  &   15 &  3  & 2,6,31 \\
    NGC 2266  &  8.80  & 0.10  &  12.95  & 0.011  &  -0.26  &   45 &  2  & 2,32 \\
    NGC 2281  &  8.70  & 0.06  &   8.92  & 0.027  &   0.13  &   35 &  3  & 1,2,6 \\
    NGC 2287  &  8.39  & 0.03  &   9.30  & 0.022  &   0.04  &   30 &  3  & 1,2,6 \\
    NGC 2301  &  8.31  & 0.03  &   9.76  & 0.023  &   0.06  &   15 &  1  & 1,2,6 \\
    NGC 2324  &  8.65  & 0.25  &  13.70  & 0.008  &  -0.40  &   70 &  8  & 2,33 \\
    NGC 2354  &  9.00  & 0.13  &  10.80  &  0.01  &  -0.30  &  100 &  5  & 2,34 \\
    NGC 2360  &  9.08  & 0.08  &  10.50  & 0.014  &  -0.15  &   25 &  4  & 1,2,35,36 \\
    NGC 2383  &  8.60  & 0.22  &  13.30  & 0.02?  & \nodata &    5 &  1  & 2,25 \\
    NGC 2395  &  9.18  & 0.07  &   8.91  & 0.02?  & \nodata &   12 &  1  & 2,8,37 \\
    NGC 2420  &  9.30  & 0.05  &  11.95  & 0.009  &  -0.32  &  140 & 12  & 2,38 \\
    NGC 2422  &  8.12  & 0.07  &   8.67  & 0.02?  & \nodata &    5 &  1  & 2,6 \\
    NGC 2437  &  8.39  & 0.15  &  11.16  & 0.023  &   0.06  &   70 &  5  & 1,2,6 \\
    NGC 2477  &  9.00  & 0.30  &  10.50  & 0.018  &  -0.05  &  190 & 28  & 2,39 \\
    NGC 2506  &  9.25  & 0.04  &  12.60  & 0.012  &  -0.20  &  130 & 12  & 2,40,41,42 \\
    NGC 2516  &  8.15  & 0.12  &   7.93  & 0.018  &  -0.05  &   35 &  6  & 2,43 \\
    NGC 2533  &  8.88  & 0.05  &  12.64  & 0.02?  & \nodata &   30 &  1  & 1,2 \\
    NGC 2539  &  8.80  & 0.06  &  10.60  & 0.018  &  -0.04  &   20 &  1  & 2,44,45 \\
    NGC 2632  &  8.90  & 0.01  &   6.39  & 0.028  &   0.14  &   30 &  5  & 1,2,6 \\
    NGC 2660  &  9.00  & 0.40  &  12.20  &  0.02  &   0.00  &  110 & 18  & 2,11 \\
    NGC 2682  &  9.60  & 0.038 &   9.65  & 0.018  &  -0.04  &  200 & 30  & 2,46\\
    NGC 2818  &  8.70  & 0.22  &  12.90  & 0.02?  & \nodata &   45 &  5  & 2,47 \\
    NGC 3114  &  8.48  & 0.07  &   9.80  & 0.02?  & \nodata &   50 &  5  & 2,48\\
    NGC 3496  &  8.78  & 0.52  &  10.70  & 0.02?  & \nodata &   70 &  4  & 2,49 \\
    NGC 3532  &  8.54  & 0.04  &   8.59  & 0.019  &  -0.02  &   90 &  9  & 2,50,51\\
    NGC 3680  &  9.27  & 0.06  &  10.20  & 0.014  &  -0.14  &   18 &  4  & 2,52\\
    NGC 3960  &  8.95  & 0.29  &  11.60  & 0.015  &  -0.12  &   50 &  4  & 2,53 \\
    NGC 4349  &  8.32  & 0.38  &  12.87  & 0.015  &  -0.12  &   25 &  1  & 1,2,6 \\
    NGC 5316  &  8.19  & 0.27  &  11.26  & 0.019  &  -0.02  &   20 &  4  & 1,2,6 \\
    NGC 5460  &  8.30  & 0.144 &   9.49  & 0.02?  & \nodata &   20 &  1  & 2,54 \\
    NGC 5617  &  8.15  & 0.54  &  11.53  & 0.02?  & \nodata &   70 &  9  & 2,55 \\
    NGC 5822  &  9.08  & 0.15  &   9.85  & 0.014  &  -0.15  &   80 &  6  & 2,56 \\
    NGC 5823  &  8.90  & 0.50  &  10.50  & 0.016  &  -0.10  &   35 &  1  & 2,6,57 \\
    NGC 6067  &  8.11  & 0.32  &  11.17  & 0.021  &   0.01  &   60 &  7  & 2,58 \\
    NGC 6208  &  9.00  & 0.18  &  10.00  & 0.019  &  -0.03  &   60 &  5  & 1,2,59 \\
    NGC 6259  &  8.34  & 0.68  &  11.50  & 0.025  &   0.10  &   85 &  3  & 1,2,60,61 \\
    NGC 6281  &  8.51  & 0.15  &   8.93  &  0.02  &   0.00  &   25 &  4  & 1,2,6 \\
    NGC 6416  &  8.78  & 0.25  &  10.12  & 0.02?  & \nodata &   35 &  3  & 2,6 \\
    NGC 6475  &  8.34  & 0.07  &   7.30  & 0.022  &   0.03  &   15 &  2  & 1,2,62 \\
    NGC 6633  &  8.80  & 0.17  &   7.80  & 0.015  &  -0.11  &   40 &  3  & 2,15 \\
    NGC 6705  &  8.40  & 0.38  &  12.70  & 0.028  &   0.14  &  110 &  1  & 1,2,6,8  \\
    NGC 6791  & 10.08  & 0.09  &  12.79  &  0.04  &   0.30  &  110 & 27  & 2,63 \\
    NGC 6802  &  8.87  & 0.85  &  10.25  & 0.007  &  -0.45  &   35 &  7  & 1,2,64 \\
    NGC 6819  &  9.40  & 0.10  &  11.80  & 0.024  &   0.07  &  270 & 33  & 2,65 \\
    NGC 6866  &  8.68  & 0.17  &  11.33  & 0.025  &   0.10  &   35 &  1  & 2,6,8 \\
    NGC 6939  &  9.11  & 0.34  &  11.30  & 0.02?  & \nodata &   80 &  4  & 2,66 \\
    NGC 6940  &  8.94  & 0.21  &  10.08  & 0.021  &   0.01  &  130 &  7  & 1,2,6 \\
    NGC 7031  &  8.14  & 0.85  &   9.77  & 0.02?  & \nodata &   15 &  1  & 1,2 \\
    NGC 7039  &  8.83  & 0.18  &  11.30  & 0.02?  & \nodata &   35 &  1  & 2,6,8 \\
    NGC 7062  &  8.70  & 0.42  &  12.76  & 0.02?  & \nodata &   35 &  2  & 2,67 \\
    NGC 7063  &  8.42  & 0.09  &   9.47  & 0.014  &  -0.16  &   15 &  1  & 2,6,8 \\
    NGC 7142  &  9.65  & 0.35  &  11.40  & 0.016  &  -0.10  &  120 & 23  & 2,68 \\
    NGC 7789  &  9.20  & 0.25  &  12.21  & 0.013  &  -0.18  &  130 & 25  & 2,69 \\
 Ruprecht 97  &  9.00  & 0.21  &  13.70  & 0.02?  & \nodata &   12 &  1  & 2,70 \\
 Ruprecht 98  &  8.78  & 0.16  &   9.42  & 0.02?  & \nodata &    7 &  1  & 2,6 \\
Ruprecht 108  &  8.41  & 0.14  &  10.21  & 0.02?  & \nodata &    5 &  1  & 2,6 \\
  Tombaugh 1  &  9.11  & 0.30  &  12.70  &  0.01  &  -0.30  &    5 &  1  & 2,71 \\
\enddata

\tablerefs{ 1: Dias et al. 2002, 2: Ahumada \& Lapasset 1995, 3:
Richtler \& Sagar 2001, 4: Carraro et al. 1994, 5: Friel \& Janes
1993, 6: Kharchenko et al. 2005, 7: Tadross 2004, 8: Loktin et al.
1994, 9: Vallenari et al. 2000, 10: Manteiga et al. 1995, 11:
Bragaglia et al. 2000, 12: Clari\'{a} et al. 2003, 13: Clari\'{a} et
al. 1994, 14: Meibom et al. 2002, 15: Hebb et al. 2004, 16: Aparicio
et al. 1991, 17: Twarog et al. 1995, 18: Sagar et al. 2001, 19: van
Leeuwen 1999, 20: Krusberg \& Chaboyer 2006, 21: Kaluzny 1988, 22:
Burke et al. 2004, 23: Subramaniam 2003, 24: Balaguer-N\'{u}\~{n}ez
et al. 2004, 25: Subramaniam \& Sagar 1999, 26: Kalirai et al. 2005,
27: Barrado y Navascu\'{e}s et al. 2001, 28: Sung \& Bessell 1999,
29: Frogel \& Twarog 1983, 30: Anthony-Twarog et al. 2005, 31:
Celeste Parisi et al. 2005, 32: Kaluzny \& Mazur 1991, 33: Piatti et
al. 2004a, 34: Clari\'{a} et al. 1999, 35: Mermilliod \& Mayor 1990,
36: Murray et al. 1988, 37: Zdanavi\v{c}jus et al. 2004, 38: Lee et
al. 2002, 39: Eigenbrod et al. 2004, 40: Carretta et al. 2004, 41:
Kim et al. 2001, 42: Marconi et al. 1997, 43: Terndrup et al. 2002,
44: Marshall et al. 2005, 45: Lapasset et al. 2000, 46: VandenBerg
\& Stetson 2004, 47: Surendiranath et al. 1990, 48: Carraro \& Patat
2001, 49: Balona \& Laney 1995, 50: Eggen 1981, 51: Sarajedini et
al. 2004, 52: Anthony-Twarog et al. 2004, 53: Bragaglia et al. 2006,
54: Barrado \& Byrne 1995, 55: Ahumada 2005, 56: Twarog et al. 1993,
57: Janes 1981, 58: Luck 1994, 59: Paunzen \& Maitzen 2001, 60:
Mermilliod et al. 2001, 61: Anthony-Twarog et al. 1989, 62: Prosser
et al. 1996, 63: Stetson et al. 2003, 64: Sirbaugh et al. 1995, 65:
Kang \& Ann 2002, 66: Andreuzzi et al. 2004, 67: Freyhammer et al.
2001, 68: Crinklaw \& Talbert 1991, 69: Barta\v{s}iut\.{e} \&
Tautvai\v{s}ien\.{e} 2004, 70: van den Bergh et al. 1976, 71: Piatti
et al. 2004b}
\end{deluxetable}

\begin{thebibliography}{}
\scriptsize \small
\bibitem[Ahumada (2005)]{ahu05} Ahumada, J. A.  2005, AN, 326, 3
\bibitem[Ahumada \& Lapasset(1995)]{ahu95} Ahumada, J., \& Lapasset, E.  1995, A\&AS, 109, 375 (AL95)
\bibitem[Andreuzzi et al. (2004)]{and04} Andreuzzi, G., Bragaglia, A., Tosi, M., Marconi, G.  2004, \mnras, 348, 297
\bibitem[Anthony-Twarog et al. (2005)]{ant05} Anthony-Twarog, B. J., Atwell, J., Twarog, B. A.  2005, \aj, 129, 872
\bibitem[Anthony-Twarog et al. (1989)]{ant89} Anthony-Twarog, B. J., Payne, D. M., Twarog, B. A.  1989, \aj, 97, 1048
\bibitem[Anthony-Twarog \& Twarog (2004)]{ant04} Anthony-Twarog, B. J., Twarog, B. A.  2004, \aj, 127, 1000
\bibitem[Aparicio et al. (1991)]{apa91} Aparicio, A., Bertelli, C., Chiosi, C., Garcia-Pelayo, J. M.  1991, ASPC, 13, 230
\bibitem[Balaguer-N\'{u}\~{n}ez et al. (2004)]{bal04} Balaguer-N\'{u}\~{n}ez, L., Jordi, C., Galad\'{\i}-Enr\'{\i}quez, D., Zhao, J. L.  2004, A\&A, 426, 819
\bibitem[Balona \& Laney (1995)]{bal95} Balona, L. A., Laney, C. D.  1995, \mnras, 277, 250
\bibitem[Barrado \& Byrne (1995)]{bar95} Barrado, D., Byrne, P. B.  1995, A\&AS, 111, 275
\bibitem[Battinelli et al.(1994)]{bat94} Battinelli, P., Brandimarti, A., Capuzzo-Dolcetta, R.  1994, A\&AS, 104, 379
\bibitem[Barta\v{s}iut\.{e} \& Tautvai\v{s}ien\.{e} (2004)]{bar04} Barta\v{s}iut\.{e}, S., Tautvai\v{s}ien\.{e}, G.  2004, Ap\&SS, 294, 225
\bibitem[Bertelli et al. (1994)]{ber94} Bertelli, G., Bressan, A., Chiosi, C., Fagotto, F., Nasi, E.  1994, A\&AS, 106, 275
\bibitem[Bica \& Alloin (1986a)]{bic86a} Bica, E., Alloin, D.  1986a, A\&AS, 66, 171
\bibitem[Bica \& Alloin (1986b)]{bic86b} Bica, E., Alloin, D.  1986b, A\&A, 162, 21
\bibitem[Bragaglia et al. (2006)]{bra06} Bragaglia, A., Tosi, M., Carretta, E., Gratton, R. G., Marconi, G., Pompei, E.  2006, \mnras, 366, 1493
\bibitem[Bragaglia et al. (2000)]{bra00} Bragaglia, A., Tosi, M., Marconi, G., Sandrelli, S.  2000, ASPC, 198, 51
\bibitem[Bressan et al. (1994)]{bre94} Bressan, A., Chiosi, C., Fagotto, F.  1994, \apjs, 94, 63
\bibitem[Bruzual \& Charlog (2003)]{bru03} Bruzual G., Charlot, S.  2003, \mnras, 344, 1000 (BC03)
\bibitem[Burke et al. (2004)]{bur04} Burke, C. J., Gaudi, B. S., DePoy, D. L., Pogge, R. W., Pinsonneault, M. H.  2004, \aj, 127, 2382
\bibitem[Carraro et al. (1994)]{car94} Carraro, G., Chiosi, C., Bressan, A., Bertelli, G.  1994, A\&AS, 103, 375
\bibitem[Carraro \& Patat (2001)]{car01} Carraro, G., Patat, F.  2001, A\&A, 379, 136
\bibitem[Carretta et al. (2004)]{car04} Carretta, E., Bragaglia, A., Gratton, R. G., Tosi, M.  2004, A\&A, 422, 951
\bibitem[Celeste Parisi et al. (2005)]{cel05} Celeste Parisi, M., Clari\'{a}, J. J., Piatti, A. E., Geisler, D.  2005, \mnras, 363, 1247
\bibitem[Chen \& Han (2004)]{che04} Chen, X. F, Han, Z. W.  2004, \mnras, 355, 1182
\bibitem[Clari\'{a} et al. (1999)]{cla99} Clari\'{a}, J. J., Mermilliod, J.-C., Piatti, A. E.  1999, A\&AS, 134, 301
\bibitem[Clari\'{a} et al. (1994)]{cla94} Clari\'{a}, J. J., Minniti, D., Piatti, A. E., Lapasset, E.  1994, \mnras, 268, 733
\bibitem[Clari\'{a} et al. (2003)]{cla03} Clari\'{a}, J. J., Piatti, A. E., Lapasset, E., Mermilliod, J.-C.  2003, A\&A, 399, 543
\bibitem[Crinklaw \& Talbert (1991)]{cri91} Crinklaw, G., Talbert, F. D.  1991, PASP, 103, 536
\bibitem[Deng et al. (1999)]{den99} Deng, L., Chen, R., Liu, X. S., Chen, J. S.  1999, \apj, 524, 824
\bibitem[Dias et al. (2002)]{dia02} Dias, W. S., Alessi, B. S., Moitinho, A., L\'{e}pine, J. R. D.  2002, A\&A, 389, 871
\bibitem[Eggen (1981)]{egg81} Eggen, O. J.  1981, \apj, 246, 817
\bibitem[Eigenbrod et al. (2004)]{eig04} Eigenbrod, A., Mermilliod, J.-C., Clari\'{a}, J. J., Andersen, J., Mayor, M.  2004, A\&A, 423, 189
\bibitem[Ferraro et al. (2003)]{fer03} Ferraro, F. R., Sills, A., Rood, R. T., Paltrinieri, B., Buonanno, R.  2003, \apj, 588, 464
\bibitem[Freyhammer et al. (2001)]{fre01} Freyhammer, L. M., Arentoft, T., Sterken, C.  2001, A\&A, 368, 580
\bibitem[Friel \& Janes (1993)]{fri93} Friel, E. D., Janes, K. A. 1993, A\&A, 267, 75
\bibitem[Frogel \& Twarog (1983)]{fro83} Frogel, J. A., Twarog, B. A.  1983, \apj, 274, 270
\bibitem[Hebb et al. (2004)]{heb04} Hebb, L., Wyse, R. F. G., Gilmore, G.  2004, \aj, 128, 2881
\bibitem[Hurley et al. (2005)]{hur05} Hurley, J. R., Pols, O. R., Aarseth, S. J., Tout, C. A.  2005, \mnras, 363, 293
\bibitem[Janes (1981)]{jan81} Janes, K. A.  1981, \aj, 86, 1210
\bibitem[Kaluzny (1988)]{kal88} Kaluzny, J.  1988, AcA, 38, 339
\bibitem[Kaluzny & Mazur (1991)]{kal91} Kaluzny, J., Mazur, B.  1991, AcA, 41, 191
\bibitem[Kang \& Ann (2002)]{kan02} Kang, Y.-W., Ann, H. B.  2002, JKAS, 35, 87
\bibitem[Kharchenko et al. (2005)]{kha05} Kharchenko, N. V., Piskunov, A. E., R\"{o}eser, S., Schilbach, E., Scholz, R.-D.  2005, A\&A, 438, 1163
\bibitem[Kim et al. (2001)]{kim01} Kim, S.-L., Chun, M.-Y., Park, B.-G., Kim, S. C., Lee, S. H., Lee, M. G., Ann, H. B., Sung, H., Jeon, Y.-B., Yuk, I.-S.  2001, AcA, 51, 49
\bibitem[Krusberg \& Chaboyer (2006)]{kur06} Krusberg, Z. A. C., Chaboyer, B.  2006, \aj, 131, 1565
\bibitem[Lapasset et al. (2000)]{lap00} Lapasset, E., Clari\'{a}, J. J., Mermilliod, J.-C.  2000, A\&A, 361, 945
\bibitem[Lee et al. (2003)]{lee03} Lee, M. G., Park, H. S., Park, J.-H., Sohn, Y.-J., Oh, S. J., Yuk, I.-S., Rey, S.-C., Lee, S.-G., Lee, Y.-W., Kim, H.-I., and 5 coauthors  2003, \aj, 126,
2840
\bibitem[Lee et al. (2002)]{lee02} Lee, S. H., Ann, H. B., Kang, Y.-W.  2002, aprm.conf, 273
\bibitem[Lejeune et al. (1997)]{lej97} Lejeune, Th., Cuisinier, F., Buser, R.  1997, A\&AS, 125, 229
\bibitem[Loktin \& Matkin (1994)]{lok94} Loktin, A. V., Matkin, N. V.  1994, A\&AT, 4, 153
\bibitem[Luck (1994)]{luc94} Luck, R. E.  1994, \apjs, 91, 309
\bibitem[Manteiga outeira et al. (1995)]{man95} Manteiga Outeira, M., Acarreta Rodriguez, J. R.,  Martinez Roger, C., Straniero, O.  1995, IAUS, 164, 377
\bibitem[Marconi et al. (1997)]{mar97} Marconi, G., Hamilton, D., Tosi, M., Bragaglia, A.  1997, \mnras, 291, 763
\bibitem[Marshall et al. (2005)]{mar05} Marshall, J. L., Burke, C. J., DePoy, D. L., Gould, A., Kollmeier, J. A.  2005, \aj, 130, 1916
\bibitem[Meibom et al. (2002)]{mei02} Meibom, S.,  Andersen, J., Nordstr\"{o}m, B.  2002, A\&A, 386, 187
\bibitem[Mermilliod et al. (2001)]{mer01} Mermilliod, J.-C., Clari\'{a}, J. J., Andersen, J., Piatti, A. E., Mayor, M.  2001, A\&A, 375, 30
\bibitem[Mermilliod \& Mayor (1990)]{mer90} Mermilliod, J.-C., Mayor, M.  1990, A\&A, 237, 61
\bibitem[Murray et al. (1988)]{mur88} Murray, R. L., Anthony-Twarog, B. J., Twarog, B. A.  1988, BAAS, 20, 717
\bibitem[Paunzen \& Maitzen (2001)]{pau01} Paunzen, E., Maitzen, H. M.  2001, A\&A, 373, 153
\bibitem[Piatti et al. (2004)]{pia04} Piatti, A. E., Clari\'{a}, J. J., Ahumada, A. V.  2004a, A\&A, 418, 979
\bibitem[Piatti et al. (2004)]{pia04} Piatti, A. E., Clari\'{a}, J. J., Ahumada, A. V.  2004b, A\&A, 421, 991
\bibitem[Piotto et al. (2002)]{pio02} Piotto, G., King, I. R., Djorgovski, S. G., Sosin, C., Zoccali, M., Saviane, I., De Angeli, F., Riello, M., Recio Blanco, A., Rich, R. M., Meylan, G., Renzini, A.  2002, A\&A, 391, 945
\bibitem[Pols \& Marinus (1994)]{pol94} Pols, O. R., Marinus, M.  1994, A\&A, 288, 475
\bibitem[Prosser et al. (1996)]{pro96} Prosser, C. F., Randich, S., Stauffer, J. R.  1996, \aj, 112, 649
\bibitem[Richtler \& Sagar (2001)]{ric01} Richtler, T., Sagar, R.  2001, BASI, 29, 53
\bibitem[Sagar et al. (2001)]{sag01} Sagar, R., Munari, U., de Boer, K. S.  2001, \mnras, 327, 23
\bibitem[Salpeter (1955)]{sal55} Salpeter, E. E.  1955, \apj, 121, 161
\bibitem[Sandage (1953)]{san53} Sandage, A. R.  1953, \aj, 58, 61
\bibitem[Sarajedini et al. (2004)]{sar04} Sarajedini, A., Brandt, K., Grocholski, A. J., Tiede, G. P.  2004, \aj, 127, 991
\bibitem[Sirbaugh et al. (1995)]{sir95} Sirbaugh, R. D., Lewis, K. A., Friel, E. D.  1995, AAS, 18710708
\bibitem[Stetson et al. (2003)]{ste03} Stetson, P. B., Bruntt, H., Grundahl, F.  2003, PASP, 115, 413
\bibitem[Stryker (1993)]{str93} Stryker, L. L.  1993, PASP, 105, 1081
\bibitem[Subramaniam (2003)]{sub03} Subramaniam, A.  2003, BASI, 31, 49
\bibitem[Subramaniam \& Sagar (1999)]{sub99} Subramaniam, A., Sagar, R.  1999, \aj, 117, 937
\bibitem[Surendiranath et al. (1990)]{Sur90} Surendiranath, R., Kameswara Rao, N., Sagar, R., Nathan, J. S., Ghosh, K. K.  1990, JApA, 11, 151
\bibitem[Tadross (2004)]{tad04} Tadross, A. L.  2004, ChJAA, 4, 67
\bibitem[Terndrup et al. (2002)]{ter02} Terndrup, D. M., Pinsonneault, M., Jeffries, R. D., Ford, A., Stauffer, J. R., Sills, A.  2002, \apj, 576, 950
\bibitem[Tian et al. (2006)]{tia06} Tian, B., Deng, L., Han, Z., Zhang, X. B.  2006, A\&A, 455, 247
\bibitem[Twarog et al. (1995)]{twa95} Twarog, B. A., Anthong-Twarog, B. J., Hawarden, T. G.  1995, PASP, 107, 1215
\bibitem[Twarog et al. (1993)]{twa93} Twarog, B. A., Anthong-Twarog, B. J., McClure, R. D.  1993, \pasp, 105, 78
\bibitem[van den Bergh et al. (1976)]{van76} van den Bergh, S., Herbst, E., Harris, G. L., Herbst, W.  1976, \apj, 208, 770
\bibitem[van Leeuwen (1999)]{van99} van Leeuwen, F.  1999, ASPC, 167, 52
\bibitem[Vallenari et al. (2000)]{val00} Vallenari, A., Carraro, G., Richichi, A.  2000, A\&A, 353, 147
\bibitem[VandenBerg \& Stetson (2004)]{van04} VandenBerg, D. A., Stetson, P. B.  2004, PASP, 116, 997
\bibitem[Xin \& Deng (2005)]{xin05} Xin, Y., Deng, L.  2005, \apj, 619, 824 (XD05)
\bibitem[Zdanavicius et al. (2004)]{zda04} Zdanavicius, J., Zdanavicius, K., Straizys, V., Kazlauskas, A., Cernis, K., Chen, C. W., Chen, W. P., Boyle, R. P., Tautvaisiene, G.  2004, Balt.A, 13, 555
\end{thebibliography}
\end{document}